\documentclass[reprint,amsmath,amssymb,pra]{revtex4-1}
\usepackage{soul}
\usepackage{graphicx}
\usepackage{dcolumn}
\usepackage{bm}
\begin{document}

\title{Bragg-induced oscillations in non-$\mathcal{PT}$ complex photonic lattices}

\author{P. A. Brand\~ao}
\email{paulocabf@gmail.com}
\author{S. B. Cavalcanti}%
\email{sbessa@gmail.com}
\affiliation{Universidade Federal de Alagoas, Cidade Universit\'{a}ria, Macei\'{o}-AL, 57072-970, Brazil}

\date{\today}

\begin{abstract}
When a monochromatic beam of light propagates through a periodic structure with the incident angle satisfying the Bragg condition, its Fourier spatial spectra oscillates between the resonant modes situated at the edges of the Brillouin zones of the lattice, exhibiting a nontrivial dynamics. Here, we investigate  these Bragg-induced oscillations in a specific complex non-$\mathcal{PT}$ periodic structure, that is, a periodic medium with gain and loss with no symmetry under the combined action of parity and time reversal operations. We compare our analytic results based on the expansion of the optical field in Bragg-resonant plane waves with a direct numerical integration of the paraxial wave equation beyond the shallow potential approximation and using a wide Gaussian beam as initial condition.  In particular, we study under which conditions a mode 
trapping phenomenon may still be observed and to inspect how the energy exchange between the spectral modes takes place during propagation in this more general class of asymmetric complex potentials. 
\end{abstract}

\pacs{42.25.Bs,42.25.Fx,42.79.Gn,}% PACS, the Physics and Astronomy
                             % Classification Scheme.
%\keywords{Suggested keywords}%Use show keys class option if keyword
                              %display desired
\maketitle

%\tableofcontents

\section{Introduction}
\label{sec1}

The unified concept of Parity-Time ($\mathcal{PT}$) symmetry introduced by Bender and Boettcher almost twenty years ago is now part of an active area of scientific research \cite{bender1998real,bender2007making,bender1999pt,bender2003must}. Systems that are invariant under the combined action of parity and time symmetries can be described by non-Hermitian operators with a real-valued spectrum. A Hamiltonian  operator is defined as $\mathcal{PT}$-symmetric if it commutes with the $\mathcal{PT}$ operator, $[\mathcal{PT},H] = 0$. However, this condition alone does not guarantee that both commuting operators will share a common set of eigenvectors because the operator $\mathcal{T}$ is antilinear. In practice, the Hamiltonian $H(b)$ contains a free parameter $b$ which may be increased 
up to a critical value so that the system undergoes a  symmetry breaking phase transition.  When the symmetry of the Hamiltonian operator is broken, the commuting property does not guarantee any longer that the Hamiltonian and the $\mathcal{PT}$ operator share a common set of eigenvectors, and the Hamiltonian spectrum becomes fully or partially complex. In the physical context, $\mathcal{PT}$ symmetry behavior has been experimentally demonstrated in optical coupled waveguides \cite{makris2008beam,guo2009observation,ruter2010observation}, silicon photonic circuits \cite{feng2011nonreciprocal}, superconducting wires \cite{rubinstein2007bifurcation} and even in classical mechanical systems \cite{bender2013observation}, to cite a few.

Even though the concept of $\mathcal{PT}$ symmetry has increased the classes of possible physical Hamiltonians by extending Hermitian systems to the complex plane, such required symmetry is not sufficient or necessary for an operator to exhibit a real-valued spectrum. Although it can be shown that the eigenvalues of a $\mathcal{PT}$-symmetric Hamiltonian always show up as complex-conjugate pairs, this property is not exclusive to the $\mathcal{PT}$-symmetric systems. A necessary and sufficient condition for the spectrum of a non-Hermitian Hamiltonian to be purely real may be formulated in terms of a more general class of pseudo-Hermitian operators. It was shown extensively by Mostafazadeh, shortly after the seminal work of Bender and Boettcher, that Hamiltonians with $\mathcal{PT}$ symmetry are actually a subgroup of a more general class of pseudo-Hermitian Hamiltonians \cite{mostafazadeh2002pseudo,mostafazadeh2002pseudo2,mostafazadeh2010pseudo,mostafazadeh2004physical}. In this theory, an operator $H$ is said to be $\eta$-pseudo-Hermitian if there exists a Hermitian invertible linear operator $\eta$ such that $H^{\dagger} = \eta H \eta^{-1}$. It was shown recently that it is possible to loosen the set of conditions on the pseudo-Hermiticity by requiring $\eta$ not to be invertible \cite{yang2017classes}. We will discuss this new class of operators in section 2 below.

Optical beam oscillations induced by Bragg-resonance were studied ten years ago by Shchesnovich and Ch\'avez-Cerda within the shallow potential approximation \cite{shchesnovich2007bragg}. Bragg-induced Rabi oscillations is an allusion to the old Rabi problem, where matter oscillations between a two-level quantum system are driven by an external optical field, so that the authors named these optical oscillations driven 
by matter as well. However, we find more transparent and unambiguous to call these Bragg-induced oscillations simply Bragg oscillations. This is because there are other types of Rabi-like oscillations present in photonic systems \cite{makris2008optical,trompeter2006bloch,PhysRevLett.102.076802}. Bloch oscillations in an optical setup with $\mathcal{PT}$ symmetry was also considered \cite{longhi2009bloch}. Until now, attention to Bragg oscillations in optical beams propagating in periodic photonic structures was devoted to linear Hermitian \cite{shchesnovich2007bragg}, nonlinear Hermitian \cite{brandao2017effect}, linear $\mathcal{PT}$-symmetric \cite{brandao2017bragg}. It should be noted here, that in this last reference within a two-mode approach, oscillations were retrieved below the symmetry breaking point while above it, depending on the initial condition, exponential growth or mode trapping. In this paper, we consider a multimode approach of Bragg oscillations in a more general class of complex photonic lattices and retrieve the oscillations, with selected modes with higher amplitudes depending on the initial mode, and we also find unidirectional energy 
transference among modes when the optical medium presents a complex refractive index.

Section 2 is devoted to a discussion of non-$\mathcal{PT}$ potential operators and their spectra, where we show explicitly how one may tailor such potentials. We also introduce the periodic potential that is used throughout the paper. Section 3 presents the evolution equations for the spectral amplitudes of individual resonant plane waves as a first-order coupled system of differential equations. This system is solved by numerical techniques in section 4 along with its physical interpretation. Our conclusions are presented in section 5.

%%%%%%%%%%%%%%%%%%%%%%%%%%%%%%%% THEORY %%%%%%%%%%%%%%%%%%%%%%%%%%%%%%%%%%%%%%
\section{Non-$\mathcal{PT}$ Hamiltonians}
\label{sec2}
As pointed out in the last section, new classes of general non-$\mathcal{PT}$ symmetric potentials with arbitrary gain and loss regions were introduced \cite{cannata1998schrodinger,miri2013supersymmetry,tsoy2014stable,nixon2016all,yang2017classes}. This is a huge step forward in the ability to construct arbitrary complex materials while still preserving a real spectrum of eigenvalues. We follow one such specific method for constructing arbitrary classes of complex materials that was introduced very recently \cite{yang2017classes}. If $L = d^{2}/dx^2 + V(x)$ is a Schr\"odinger operator and if there exists another operator $\eta$ such that $L$ and $\eta$ are related by a similarity relation
\begin{equation}
    \eta L = L^{*}\eta,
    \label{condition}
\end{equation}
then it is easy to show that all eigenvalues of $L$ appear in conjugate pairs if the kernel of $\eta$ is empty \cite{nixon2016all}. This resembles the approach developed by Mostafazadeh \cite{mostafazadeh2002pseudo,mostafazadeh2002pseudo2} but, in the present situation, $\eta$ need not be invertible. Given the generality of the $\eta$ operator, we choose it to be a combination of the parity operator $\mathcal{P}$ and differential operators as in
\begin{equation}
    \eta = \mathcal{P}\left[ \frac{d}{dx} + h(x) \right],
\end{equation}
where $h(x)$ is a complex function whose properties we will now derive. When the right hand side of \eqref{condition} acts on an arbitrary function $g(x)$ we have
\begin{equation}
\begin{split}
%L^{*}\eta g(x) & = \left[ \frac{d^2}{dx^2} + V^{*}(x)\right]\mathcal{P}\left[ \frac{dg(x)}{dx} + h(x)g(x) \right] \\
% & = \left[ \frac{d^2}{dx^2} + V^{*}(x)\right]\left[ -\frac{dg(-x)}{dx} + h(-x)g(-x) \right] \\
% & =  -\frac{d^{3}g(-x)}{dx^3} + \frac{d^{2}[h(-x)g(-x)]}{dx^2} - V^{*}(x)\frac{dg(-x)}{dx}  + V^{*}(x)h(-x)g(-x) ,
 L^{*}\eta g(x) & =  -\frac{d^{3}g(-x)}{dx^3} + \frac{d^{2}[h(-x)g(-x)]}{dx^2} \\
                & - V^{*}(x)\frac{dg(-x)}{dx}  + V^{*}(x)h(-x)g(-x) ,
\end{split}
\end{equation}
while the left hand side of \eqref{condition} reads
\begin{equation}
\begin{split}
    \eta L g(x) & =   -\frac{d^{3}g(-x)}{dx^{3}} - \frac{d[V(-x)g(-x)]}{dx} \\
                & + h(-x)\frac{d^{2}g(-x)}{dx^2} + h(-x)V(-x)g(-x). \\
\end{split}
\end{equation}
Since \eqref{condition} must be satisfied, after letting $x\rightarrow -x$ and setting all coefficients of $g(x)$ and $dg(x)/dx$ equal to zero (because $\eta L - L^{*}\eta$ is the null operator), we have the following coupled system of differential equations
\begin{equation}
    V(x)-V^{*}(-x) = 2\frac{dh(x)}{dx},
    \label{coupled1}
\end{equation}
and
\begin{equation}
    [V(x)-V^{*}(-x)]h(x) = \frac{d^{2}h(x)}{dx^{2}} - \frac{dV(x)}{dx}.
    \label{coupled2}
\end{equation}
After taking the complex conjugate of \eqref{coupled1} together with the substitution $x\rightarrow -x$, it is easy to see that
\begin{equation}
    -\frac{dh^{*}(-x)}{dx} = \frac{1}{2}[V^{*}(-x)  - V(x)] = -\frac{dh(x)}{dx},
\end{equation}
which follows that $dh^{*}(-x)/dx = dh(x)/dx$ and, after integrating both sides, $h^{*}(-x) = h(x) + c_{1}$ with $c_{1}$ being the constant of integration. Next, let us substitute the right hand side of \eqref{coupled1} into the left hand side of \eqref{coupled2} to obtain
\begin{equation}
    2h(x)\frac{dh(x)}{dx} = \frac{d^{2}h(x)}{dx^{2}} - \frac{dV(x)}{dx},
\end{equation}
and, after integrating once,
\begin{equation}
    \int^{h(x)}2hdh = \int^{\frac{dh}{dx}} d\left( \frac{dh}{dx'} \right) - \int^{V(x)}dV + c_{2},
\end{equation}
where $c_{2}$ groups all constants of integration. It is then easy to see that $V(x)$ is given by
\begin{equation}
    V(x) = \frac{dh(x)}{dx} - h^{2}(x) + c_{2}.
    \label{potentialc2}
\end{equation}
Since $c_{2}$ only adds a constant value to the potential function, we may, without loss of generality, choose it as $c_{2} = 0$. To find the value of $c_{1}$ we substitute \eqref{potentialc2} and the expression $h^{*}(-x) = h(x)+c_1$ into \eqref{coupled1}
\begin{equation}
    c_{1}[2h(x) + c_{1}] = 0.
\end{equation}
Therefore, we can choose $c_{1} = 0$. Since now $h^{*}(-x) = h(x)$ it follows that the function $h(x)$ is $\mathcal{PT}$-symmetric. So, by choosing a $\mathcal{PT}$-invariant function $h(x)$  one may generate the potential function 
\begin{equation}
    V(x) = \frac{dh(x)}{dx} - h^{2}(x).
\end{equation}
The choice of the $\mathcal{PT}$-symmetric $h(x)$ function is completely arbitrary and, to study a periodic potential, we follow \cite{yang2017classes} in choosing the convenient expression $h(x) = 1 + \cos x + ib\sin x$, with $b$ as a free real and positive parameter. The potential is then given explicitly by
\begin{multline}
    V(x)  = b^{2}\sin^{2}x - \cos^{2}x - \sin x - 2\cos x -1 \\
          + ib[\cos x - 2\sin x - \sin(2x)],
    \label{potentialir}
\end{multline}
that is, a potential complex function whose real and imaginary parts are not even and odd respectively, and it is a periodic function with period $2\pi$. Nevertheless, the spectrum of \eqref{potentialir} is completely real for $b<1$ and partially complex for $b>1$ as shown in \cite{yang2017classes}. Therefore, this is an explicit example of a potential that is not $\mathcal{PT}$-symmetric and yet has a phase transition point. In the next section, we will address the problem of light propagation through an asymmetric potential function as given in \eqref{potentialir}. In particular, we will be interested in the evolution of the Bragg modes when the lattice is at the breaking point $b = 1$.

%In a PT-symmetric system, at the breaking point real degenerated eigenvalues split into a conjugate pair. While in
%an asymmetric potential, the appearance of the complex pair stems from
%the collision of two isolated real eigenvalues. In the next section, we will address the problem of light propagation through an asymmetric potential function as given in \eqref{potentialir}. In particular, we will be interested in the evolution of the Bragg modes when the lattice is at the breaking point $b = 1$.

\section{Evolution equations}
We will assume from now on that the propagation of monochromatic scalar optical beams $\psi(x,z)$ is well described by the dimensionless (1+1)-dimensional paraxial wave equation
\begin{equation}
i\psi_{z}= - \psi_{xx} + V(x)\psi,
\label{paraxial}
\end{equation}
where $z$ is the propagation distance, $x$ the transverse coordinate and $V(x)$ is the potential \eqref{potentialir} representing the properties of the photonic lattice. Before we discuss the ansatz used to solve \eqref{paraxial} we decompose the potential $V(x)$ into its harmonic components
\begin{equation}
    V(x) = \sum_{n=-\infty}^{\infty}V_{n}e^{inx} %= V_{-2}e^{-2ix} + V_{-1}e^{-ix} + V_{0} + V_{1}e^{ix} + V_{2}e^{2ix},
    \label{potential}
\end{equation}
where the, generally complex, Fourier amplitudes of the potential are given by $V_{-2} = -(1/4)(1-b)^{2}$, $V_{-1}=-(i/2+1)(1-b)$, $V_{0} = -[1+(1-b^{2})/2]$, $V_{1} = (i/2-1)(1+b)$ and $V_{2} = -(1/4)(1+b)^2$. The fact that some Fourier amplitudes are complex numbers is a clear signature of the non-$\mathcal{PT}$ nature of the potential considering that, for  $\mathcal{PT}$-symmetric potentials, $V(x) = \sum_{n}V_{n}e^{inx} = \sum_{n}V_{n}^{*}e^{inx} = V^{*}(-x)$, and thus $V_{n} \in \mathbb{R}$. 

Since it is known that only resonant Bragg modes are coupled during propagation when the incident beam is wide \cite{shchesnovich2007bragg,brandao2017bragg,brandao2017effect}, we choose the ansatz to be
\begin{equation}
    \psi(x,z) = \sum_{n = 0,\pm 1,\pm 2, ...}\psi_{n}(z)\exp(inx),
    \label{ansatz}
\end{equation}
where $\psi_{n}(z)$ represents the spectral amplitude of mode $n$ at a distance $z$. After substituting expressions \eqref{potential} and \eqref{ansatz} into the paraxial wave equation \eqref{paraxial} we arrive at the following set of coupled first order differential equations for the spectral amplitudes $\psi_{n}(z)$:
\begin{multline}
    i\frac{d\psi_{n}}{dz} = (n^2 + V_{0})\psi_n + V_{-2}\psi_{n+2} \\
    + V_{-1}\psi_{n+1} + V_{1}\psi_{n-1} + V_{2}\psi_{n-2}.
    \label{coupled}
\end{multline}
We will also be interested in the power as a function of $z$, which can be calculated directly from the field $\psi(x,z)$ or as an incoherent sum of all spectral amplitudes $\psi_{n}(z)$:
\begin{equation}
    P(z) = \int_{-\infty}^{\infty}|\psi(x,z)|^{2}dx = \sum_{n = 0,\pm 1,\pm 2, ...}|\psi_{n}(z)|^{2}.
    \label{power}
\end{equation}
The second equal sign used in relation \eqref{power} is not strictly correct since plane waves have infinite energy, so the right-hand side of this equation should be viewed as energy per unit (transverse) length. The general outline for the rest of the paper is to solve system \eqref{coupled}, by considering a finite number of modes, subjected to initial conditions and to determine its power and Fourier spectra evolution.

\section{Numerical results and discussion}

\begin{figure}[ht]
\includegraphics[width=7cm]{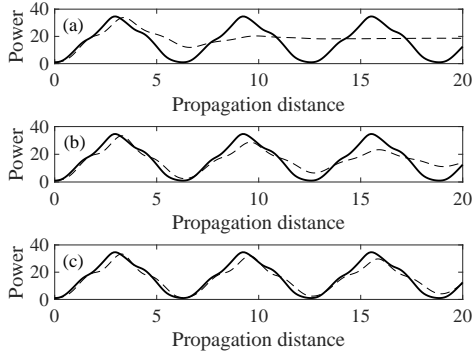}
\centering
\caption{Power evolution in a non-$\mathcal{PT}$- symmetrical complex lattice defined in \eqref{potentialir}. Continuous lines represent the solutions of \eqref{coupled}. Dashed lines are plotted from the numerical solution of the paraxial wave equation with a Gaussian input of the form $\psi(x,0)=\exp[-(x/W)^{2}]$ with (a) $W = 10$, (b) $W = 25$ and (c) $W = 50$.}
\label{fig1}
\end{figure}

\begin{figure}[ht]
\includegraphics[width=7cm]{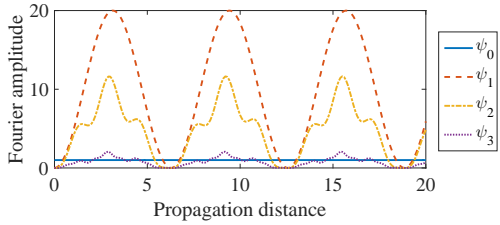}
\centering
\caption{Evolution of the Fourier amplitudes $|\psi_{n}(z)|^{2}$ for $n = 0,1,2$ and 3, corresponding to the parameters of Figure 1.}
\label{fig2}
\end{figure}

\begin{figure}[ht]
\includegraphics[width=8cm]{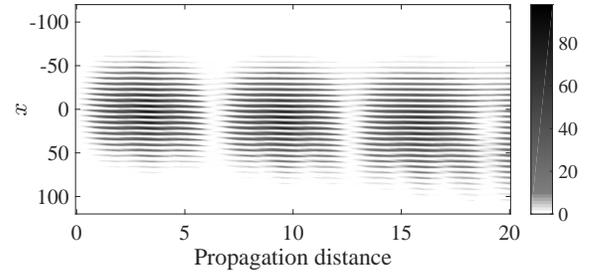}
\centering
\caption{Intensity evolution in the $(x,z)$ plane for a Gaussian beam propagating through the lattice \eqref{potentialir} with $W = 50$. The regions where the intensity is a minimum are related by the power evolution of Figure 1.}
\label{fig3}
\end{figure}

Let us assume that only eleven modes are coupled through relation \eqref{coupled}: $n\in\{\pm5,\pm4,\pm3,\pm 2,\pm1,0\}$. Within this approximation, the system \eqref{coupled} may 
be solved numerically after specifying the initial conditions for the eleven spectral amplitudes $\psi_{n}(0)$. Let us consider 
the lattice at the symmetry breaking point, $b = 1$, and that only mode $n = 0$ is initially populated, $\psi_{0}(0) = 1$, 
with all others modes, $\psi_{n}(0) = 0$ . Figure 1 shows the power evolution, $P(z) = \sum_{n}|\psi_{n}(z)|^{2}$, calculated in terms of the spectral amplitudes, along with the power evolution, $P(z) = \int|\psi(x,z)|^{2}dx$, calculated from the numerical 
solution of \eqref{paraxial} with a Gaussian beam as initial condition, given by
\begin{equation}
\psi(x,0) = \exp\left[ -\frac{1}{2}\left( \frac{x}{W} \right)^{2} \right],
\label{gaussian}
\end{equation}
where $W$ is the initial beam width. Part (a) of Figure \ref{fig1} depicts the power evolution with $W = 10$ and parts (b) and (c) with $W = 25$ and $W = 50$, respectively. One may conclude from Figure \ref{fig1} that our approximation describes better the power evolution for wider initial beams and therefore one should expect this finite-width beam approach to 
corroborate the results obtained by the  plane wave approach. Also, by closely inspecting the profiles 
shown in Figure \ref{fig1}, one may note that the power evolution is not exactly a sine or a cosine 
function due to the appearance of some 
relatively small bumps between the maximum and minimum points. Until now, to the best of our 
knowledge, no explanation or study has been given to the physical mechanism behind such specific 
behavior of the power evolution. However, this behavior may  be understood, by examining the 
dynamics involved in the excitation of the 
Bragg-modes of the wavefield, as we now show. 
Figure 2 depicts the excited Fourier modes \{$\psi_{0}$, $\psi_{1}$, $\psi_{2}$, $\psi_{3}$\} during propagation. The modes with $n$ negative 
are exactly zero for this particular set 
of parameters and the modes \{$\psi_{4}$, $\psi_{5}$\} 
are negligible on the scale that is shown 
in Figure \ref{fig1} and are therefore suppressed. 
Surprisingly, mode $\psi_{0}$ is trapped at $k=0$, 
in the sense that its spectral energy does not 
change during propagation. On the other hand, 
the modes $\psi_{1}$, $\psi_{2}$ and $\psi_{3}$ 
experience power 
oscillations \cite{shchesnovich2007bragg,
brandao2017bragg}. One 
may now conclude, that the small bumps 
appearing in Figure \ref{fig1} are mainly due to the 
excited Bragg mode $\psi_{2}$, as shown 
in Figure \ref{fig2}.
One may note that 
the sum of the four curves in this plot is equal 
to the resultant continuous curve for the 
power evolution shown in Figure 1. This is 
to be expected according to relation \eqref{power}.
Since only modes with $n\geq 0$ are excited 
in this scenario, we expect the beam to propagate 
in the positive direction of the $x$ axis. This is 
seen in Figure \ref{fig3} where the 
intensity, $|\psi(x,z)|^{2}$, of the Gaussian 
beam with 
$W = 50$ is shown. By comparing Figures 
\ref{fig1} and \ref{fig3}, it can be seen 
that the points of minimum value in the power 
evolution can be correlated to the minimum 
values of the intensity, as expected.

\begin{figure}[ht]
\includegraphics[width=6cm]{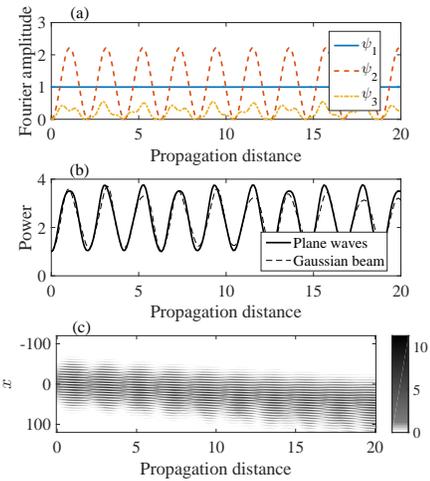}
\centering
\caption{(a) Spectral amplitudes evolution with $\psi_{1}(0) = 1$ and all other $\psi_{n}(0)=0$  of the lattice corresponding to Figures 1-3. (b) Power evolution calculated from the spectral amplitudes (solid line) and from the Gaussian beam (dashed line). (c) Intensity in the $(x,z)$ plane for the Gaussian beam with $W = 50$.}
\label{fig4}
\end{figure}

Next, we consider the spectral component $\psi_1$ to be initially populated, $\psi_{1}(0) = 1$, and all others empty. The evolution of the spectral amplitudes $\psi_{1}$, $\psi_{2}$ and $\psi_{3}$ is shown in Figure \ref{fig4}(a), after numerically solving the system \eqref{coupled}. The other eight spectral components are either zero or negligible on the scale of the plot and are, therefore, not shown. Once more, the initially populated amplitude is trapped during propagation. On the other hand, the Bragg modes $\psi_2$ and $\psi_3$ undergo periodic Rabi-like oscillations and are therefore continuously exchanging energy with the medium. The amplitude of oscillation of mode $\psi_2$ is approximately two times the amplitude of mode $\psi_1$. Part (b) of Figure 4 depicts the power evolution for these parameters illustrating a very close agreement between the model \eqref{ansatz} represented by the continuous line, and the simulation of an initial Gaussian beam $\psi(x,0) = \exp\left[ -(1/2)\left( x/W \right)^{2} \right]\exp(ix)$ centered at $k=1$, represented by the dashed line. The power oscillates in a more sinusoidal form than the previous situation presented in Figure 1. 
Figure \ref{fig4}(c) depicts the intensity evolution $|\psi(x,z)|^{2}$ for $W = 50$. As only positive Bragg modes are excited, 
the beam has a tendency to travel to the right, as expected. 
Comparing the intensity evolution displayed in part (c) with the 
power evolution in part (b), one finds that the beam exchanges much less energy from the medium compared to Figure \ref{fig1}. This is due to the fact that the spectral component $\psi_2$ oscillates with a much smaller amplitude than the spectral component $\psi_1$ as one may note from Figure \ref{fig2}.

\begin{figure}[ht]
\includegraphics[width=6cm]{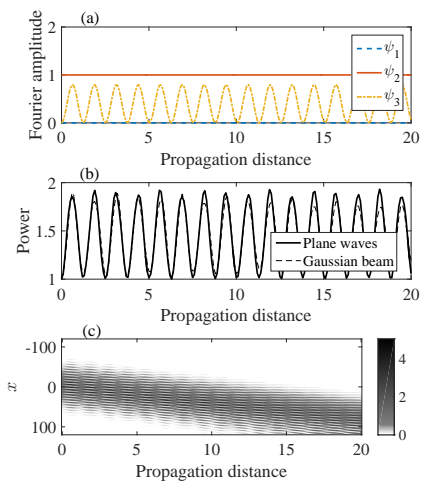}
\centering
\caption{(a) Spectral amplitudes evolution with $\psi_{2}(0) = 1$ and 
all other $\psi_{n}(0)=0$,  of the lattice corresponding to Figures 1-3. 
(b) Power 
evolution calculated from the spectral amplitudes (solid line) and from 
the Gaussian beam (dashed line). (c) Intensity in the $(x,z)$ plane for 
the Gaussian beam with $W = 50$.}
\label{fig5}
\end{figure}

Now let us address one of the situations considered in \cite{yang2017classes} where the mode $\psi_2$ is initially populated. Figure \ref{fig5} shows the results of the numerical simulations involving plane waves and a Gaussian beam given by $\psi(x,0) = \exp\left[ -(1/2)\left( x/W \right)^{2} \right]\exp(2ix)$ to match the initially populated mode $\psi_2$. Once more, the energy of the initially populated spectral mode is trapped, as can be seen in part (a) of Figure \ref{fig5}. But in this situation, besides the $\psi_2$, only mode $\psi_3$ contributes to the evolution of the beam by oscillating Rabi-like during propagation, as the yellow color line in Figure \ref{fig5}(a) shows. From 
previous discussions, we conclude that 
the Bragg contributions to the power evolution are given by a constant function and an oscillatory one with peculiar features stemming from the simultaneous excitation of various modes: the initial input mode (say mode $\psi_n$), for which $\psi_{n}(0) = 1$ becomes trapped with constant amplitude while the right-handed neighboring mode $\psi_{n+1}$, for which $\psi_{n+1}(0) = 0$
is supplied with
a larger portion of energy from matter.   Part (b) of Figure \ref{fig5} shows the power evolution from the Bragg modes (continuous line) compared to the Gaussian beam power (dashed line). The agreement between the two approaches seems to be very good. Since only two modes are excited during propagation, $\psi_2$ and $\psi_3$, we expect the beam to propagate mainly toward
the positive $x$ direction. This can be verified by inspecting Figure \ref{fig5}(c) where the intensity evolution for a Gaussian beam with $W = 50$ is shown and is clearly unidirectional.

\begin{figure}[ht]
\includegraphics[width=6cm]{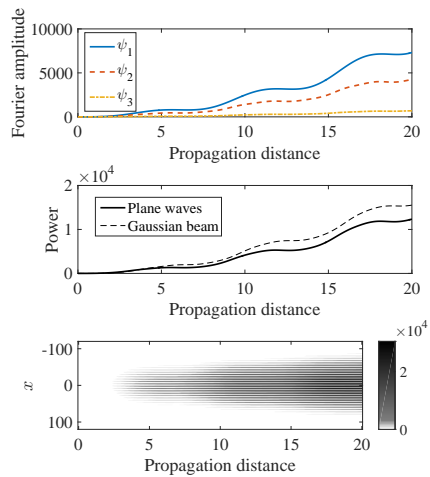}
\centering
\caption{(a) Spectral amplitudes evolution with $\psi_{-1}(0) = 1$ and $\psi_{n} (0) = 0$ for the lattice corresponding to Figures 1-3. (b) Power evolution calculated from the spectral amplitudes (solid line) and from the Gaussian beam (dashed line). (c) Intensity in the $(x,z)$ plane for the Gaussian beam with $W = 50$.}
\label{fig6}
\end{figure}

\begin{figure}[ht]
\includegraphics[width=7cm]{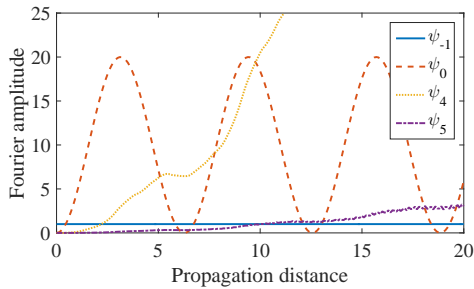}
\centering
\caption{Fourier amplitude evolution for four modes corresponding to the system in Figure \ref{fig6}. One can clearly see that mode $\psi_{0}(z)$ is stable and oscillates in a Rabi-like fashion even though the beam is incident in the negative $x$ direction.}
\label{fig7}
\end{figure}

The existence of a pure trapped Bragg mode, as predicted in \cite{brandao2017bragg} for $\mathcal{PT}$-symmetric lattices seems to be very rare and probably only achievable within the shallow potential approximation. So, even though the intensity pattern depicted in Figure \ref{fig5}(c) suggests a transparent medium, there is another excited Bragg mode and, therefore, visible to the lattice, which reflects the field even with the lattice at the symmetry breaking point $b = 1$. We are now in a position to discuss an incident beam that excites Bragg modes with $n$ negative. However, as demonstrated for $\mathcal{PT}$-symmetric 
lattices \cite{brandao2017bragg}, when the photonic lattice is at the symmetry breaking point and the initial input power is 
totally in $k= -1$,
the Bragg modes continuously absorb energy and diverge during propagation. Therefore, we expect our eleven-modes approach to become invalid for 
this regime, at least when $z$ is large. 
To verify these claims, we present in Figure \ref{fig6} the spectral evolution 
for a wavefield whose spectral amplitude $\psi_{-1}(0) = 1$ is 
initially populated while all others are zero. We clearly see that the amplitudes \{$\psi_{1}$,$\psi_{2}$,$\psi_{3}$\} 
grow indefinitely compared to the others. This 
implies that the medium is continuously 
giving energy to the field which becomes 
unbounded in amplitude. Figure \ref{fig6}(b) 
shows the power evolution predicted by 
\eqref{coupled} and we conclude that it 
describes very well the power evolution when compared to a Gaussian beam $\psi(x,0) = \exp\left[ -(1/2)\left( x/W \right)^{2} \right]\exp(-ix)$ 
with $W = 50$. Some other nonzero spectral 
amplitudes are shown in Figure \ref{fig7} where, surprisingly, 
we see that the spectral amplitude $\psi_{0}$ does not grow indefinitely during propagation. It reproduces a pure Rabi-like profile exchanging energy periodically with the medium. This 
is a remarkable result as one finds
that a stable Fourier amplitude oscillation is achieved in this unusual configuration, that is, at the critical point of a complex non-$\mathcal{PT}$ 
optical lattice. Note that the coupling, in all cases, depends strongly on the initial conditions:
the initial mode, say $\psi_n$, always become trapped while the mode $\psi_{n+1}$ becomes strongly coupled as in the previous 
cases. 

\section{Conclusions}
We have studied Bragg oscillations in a non-$\mathcal{PT}$-symmetric complex photonic lattice. We find rich dynamics with asymmetric energy exchange between Bragg modes during the propagation of an optical field. Depending on the incident angle, Rabi-like oscillations are indeed achievable in complex photonic lattices, at the critical point. Power oscillations exhibit peculiar features stemming from the excitation of several modes, besides the initial one. The trapping of a pure Bragg mode with a non-vanishing amplitude  in the more general context of a lattice was not observable. We have also found  evidence of stable Fourier spectra evolution even 
when the incident input power is concentrated in $\psi_n$ with $n < 0$ , which means, a beam travelling in the negative $x$ direction. These are unusual features: a stable Fourier amplitude evolution that persists even with negative $n$ and particularly at the critical point, where the Hamiltonian spectrum becomes partially complex. It should be noted here that, in spite of the simple model used here, we have been able to compare its results with numerical ones and shown that the former are quite reliable.  Due to its relative simplicity, we hope that the present model might offer insights into the study of systems described by a complex photonic lattice described by a non-Hermitian complex index of refraction. The extension of optical systems to the complex plane should well open up the way to a new generation of 
optical devices and techniques useful in the pursue of the control of light.

%\bibliography{sample}

\end{document}